\documentclass[aps,twocolumn,showpacs,eqsecnum,amsmath,amssymb]{revtex4}

\usepackage{color}
\usepackage{graphicx}
\usepackage{dcolumn}
\usepackage{bm}
\usepackage{amsthm}
\usepackage{amsthm}
\newtheorem{theorem}{Theorem}[section]

\theoremstyle{definition}

\theoremstyle{remark}

\begin{document}

\title{Spectra and entropy of multi-field warm inflation}

\author{Xi-Bin Li}
\email{lxbbnu@mail.bnu.edu.cn}
\affiliation{Department of Physics, Beijing Normal University, Beijing 100875, China}
\affiliation{Department of Astronomy, Beijing Normal University, Beijing 100875, China}
\affiliation{School of Physics and Technology, Wuhan University, Wuhan 430072, China}

\author{Xiao-Gang Zheng}
\email{xiaogang.zheng@whu.edu.cn}
\affiliation{School of Physics and Technology, Wuhan University, Wuhan 430072, China}

\author{Jian-Yang Zhu}
\thanks{Corresponding author}
\email{zhujy@bnu.edu.cn}
\affiliation{Department of Physics, Beijing Normal University, Beijing 100875, China}

\date{\today}

\begin{abstract}

We study the power spectra and entropy of two-field warm inflationary scenario with canonical condition which is described by many-dimensional stochastic differential equations. The field perturbations are analytically calculated via a Volterra integral equation of the second kind, based on which we obtain a spectra with leading order and first order of slow-roll parameters. We also find the evolutions of background are not independent but relying on dissipative coefficients, which is distinguished from that in cold inflation. Then, we calculate the entropy on the basis of  statistical physics theory by introducing an entropy matrix. On super-horizon scale, the entropy matrix follows the fluctuation-dissipation relation consistent with the scale-invariance of spectra or the stationarity of field perturbations. The entropy perturbation vanishes at both super-horizon and sub-horizon scale, while narrow peaks generate at a specific scale which could be considered as horizon. In addition, the second law of thermodynamics is followed as well.
\end{abstract}
\maketitle

\section{\label{introduction}Introduction}

Warm inflation model was established as a candidate scenario to overcome some defects in cold inflation \cite{doi:10.1142/S0217751X09044206,BARTRUM2014116}.
However, it was realized a few years after its original proposal that the idea of warm inflation was not easy to be realized in concrete models and even is simply not possible  \cite{PhysRevD.60.083509,PhysRevLett.83.264}. Some problems were mentioned in such a scenario. Shortly afterwards, many successful models of warm inflation were established, in which improved method is based on the theoretical model that the inflaton indirectly interacts with the light degrees of freedom through a heavy mediator fields instead of being coupled with a light field directly \cite{PhysRevD.84.103503,PhysRevLett.117.151301,1475-7516-2011-09-033}.
In such a scenario, the evolution of inflaton field can be properly determined in the context of the in-in form, or Schwinger closed-time path functional formalism \cite{rammer2007quantum}. This formalism leads an stochastic differential equation which contains a dissipative term and a Gaussian stochastic noise term as a type of generalized Langevin-like equation of motion \cite{PhysRevD.76.083520,PhysRevD.91.083540}.

Recently, series of work has been done to distinguish warm inflationary scenario from the cold one. The warm little inflation scenario can lead to different realizations of warm inflation on being both dynamically
and observationally consistent \cite{PhysRevD.98.083502}. Within the warm inflationary scenario, the modifications to the primordial perturbation spectrum induced by dissipative and thermal effects generically lead to a more blue-tilted scalar spectrum with respect to cold inflation models with the same potential functions. A more suppressed tensor component is also obtained in previous work \cite{PhysRevD.97.063516} compared with cold inflation which is another method to distinguish the two models. Besides, the warm inflation model with appropriate dissipative coefficient, like $\Gamma(\phi)=\Gamma_2 \phi^2$, dramatically increases the possibility of the occurrence of inflation \cite{PhysRevD.98.043510}.

Compared with the predictions of cold inflation that primordial density fluctuations mostly from quantum fluctuation and thermal bath are only generated at the end of inflation \cite{RevModPhys.78.537}, warm inflation model suggests that our Universe is hot during the whole inflation when inflaton fields couple with the thermal bath and the primary source of density fluctuations comes from thermal fluctuations \cite{PhysRevD.64.063513, PhysRevD.62.083517, PhysRevD.69.083525}. The equation of motion for warm inflation can be written as a stochastic Lengevin equation, in which there is a dissipation term to describe the inflaton fields coupling with thermal bath and there is also a fluctuation term described by a stochastic noise term \cite{PhysRevD.71.023513, PhysRevD.76.083520}. The fundamental principles of warm inflation have been reviewed recently in Ref. \cite{PhysRevD.50.2441}.

Besides single-field inflationary scenario, a different possible way to generate perturbations in agreement with observations is so-called multi-field inflationary model. Multi-field inflationary scenarios usually involves several fields which play a dynamical role during inflation \cite{PhysRevD.53.5437, PhysRevD.54.7181, MUKHANOV199852, PhysRevD.59.123512, STAROBINSKY2001383, PhysRevD.64.083514, 0264-9381-21-2-002, PhysRevD.96.103533}. Previously, two-field inflation with canonical kinetic term was investigated by decomposing field perturbations into perturbations parallel to background trajectory in field space (the adiabatic or curvature perturbation) and orthogonal to the background trajectory (the isocurvature or entropy perturbation) \cite{PhysRevD.63.023506}. Recently, such a model has been successfully generalized to the warm inflationary scenario \cite{PhysRevD.97.063510}.

In previous works, relevant results have been obtained under noncanonical situation \cite{PhysRevD.67.063512, 1475-7516-2007-07-014, 1475-7516-2009-04-020}, but in this paper we only concentrate on the canonical situation since it allows analytical expressions. We will continue to focus on multi-field warm inflation but some areas have been improved. First, the field perturbations are obtained by a description of many-dimensional stochastic differential equations which is equivalent to an integral equation set. Second, we study the entropy perturbation via stochastic physical method by which we introduce a symmetric and non-negative matrix.

The organization of this paper is the following. In Sec. \ref{spectra}, from a phenomenological stochastic differential equation set which describe the evolution of field perturbations, a scale-invariant spectra is derived. Sec. \ref{entropy_warm_inflation} is devoted to the study of entropy on both super-horizon and sub-horizon scale based on which discussions on relevant properties have also been made. Finally, in Sec. \ref{conclusion}, we conclude our work and give some further discussions about our results.

\section{\label{spectra}Spectra of multi-fields warm inflation}                                       

In warm inflationary scenario, a inflaton $\Phi(\textbf{x},t)$ is composed of a unperturbed background field $\phi(t)$ and a perturbed field $\delta\phi(\textbf(x),t)$ which follow the equations
\begin{eqnarray}
     \frac{\partial^2 \phi}{\partial t^2}+\left[3H+\Upsilon\right]\frac{\partial \phi}{\partial t}+V,_\phi(\phi) &=& 0,\label{EOM_background_inflaton} \\
    \Big\{\frac{\partial^2}{\partial t^2}+[3H+\Upsilon(\phi)]\frac{\partial}{\partial t}-\frac{1}{a^2}\nabla^2+\nonumber \\
    \Upsilon_\phi(\phi)\dot{\phi}
    +V_{\phi\phi}(\phi)\Big\}\delta\varphi &=& \xi_T, \label{EOM_pertubed_inflaton}
\end{eqnarray}
where $\Upsilon$ is the dissipation coefficient and $\xi_T$ is the thermal noise fluctuation. In this paper, we consider only the case of de Sitter space-time, where $a(t)=\textrm{exp}(Ht)$ and $H$ is regarded as a constant. According to the fluctuation-dissipation theorem, dissipation coefficient $\Upsilon$ and fluctuation noise $\xi_T$ follow the relation
\begin{eqnarray}
    \langle \xi_T(\textbf{x},t)\xi_T^*(\textbf{x},t')\rangle=2\Upsilon T a^{-3}\delta^3(\textbf{x}-\textbf{x}')\delta(t-t'). \label{dis_flu_relation}
\end{eqnarray}
The Fourier transformation of Eq.~(\ref{dis_flu_relation}) is
\begin{eqnarray}
    \langle \xi_T(\textbf{k},t)\xi_T^*(\textbf{k}',t')\rangle=2(2\pi^3)\Upsilon T a^{-3}\delta^3(\textbf{k}-\textbf{k}')\delta(t-t'). \nonumber\\ \label{dis_flu_relation_k}
\end{eqnarray}
Now extend the single field warm inflationary model to multi-field condition which is described by a phenomenologically multi-dimensional Langevin equation. The background fields follow the equations
\begin{subequations}\label{EOM_background}
\begin{eqnarray}
     \frac{\partial^2 \phi}{\partial t^2}+[3H+\Upsilon_\phi(\phi,\chi)]\frac{\partial \phi}{\partial t}+V,_\phi(\phi,\chi) &=& 0,  \\ \label{EOM_background_phi}
     \frac{\partial^2 \chi}{\partial t^2}+[3H+\Upsilon_\chi(\phi,\chi)]\frac{\partial \chi}{\partial t}+V,_\chi(\phi,\chi) &=& 0,     \label{EOM_background_chi}
\end{eqnarray}
\end{subequations}
where $V(\phi,\chi)=V_1(\phi)+V_2(\chi)+V_I(\phi,\chi)$ is potential function. The perturbed fields follow the Langevin equations
\begin{subequations}\label{EOM_pertubed}
\begin{eqnarray}
    \Big\{\frac{\partial^2}{\partial t^2}+[3H+\Upsilon_\phi]\frac{\partial}{\partial t}+\frac{k^2}{a^2}+\Upsilon_{\phi,\phi}\dot{\phi}
    +V,_{\phi\phi}\Big\}\delta\phi\nonumber \\
    +V,_{\phi\chi}\delta\chi+\Upsilon_{\phi,\chi}\dot{\phi}\delta\chi = \xi_\phi, \label{EOM_pertubed_phi} \\
    \Big\{\frac{\partial^2}{\partial t^2}+[3H+\Upsilon_\chi]\frac{\partial}{\partial t}+\frac{k^2}{a^2}+\Upsilon_{\chi,\chi}\dot{\chi}
    +V,_{\chi\chi}\Big\}\delta\chi\nonumber \\
    +V,_{\phi\chi}\delta\phi+\Upsilon_{\chi,\phi}\dot{\chi}\delta\phi = \xi_\chi. \label{EOM_pertubed_chi}
\end{eqnarray}
\end{subequations}
$\xi_\phi$ and $\xi_\chi$ are Gaussian fluctuating forces which also follow the fluctuation-dissipation relation and $,_\phi$ denotes the partial derivative with respect to $\phi$. To be convenient, we remark the symbols as a recognizable way: the physical symbols concerned with $\delta\phi$ or $\phi$ are labelled as 1 while $\delta\chi$ or $\chi$ are labelled as 2. For example,
$\boldsymbol{\varphi}=\delta\phi_i=(\delta\phi,\delta\chi)^T\equiv(\delta\phi_1,\delta\phi_2)^T$, or $\boldsymbol{\phi}=\phi_i=(\phi,\chi)^T=(\phi_1,\phi_2)^T$, where operator $T$ represents a transpose operation. Based on the slow-roll approximation,
write the background fields in Eq.~(\ref{EOM_background}) as the form
\begin{eqnarray}
    3H(1+r_i)\dot{\phi}_i+V,_i(\phi_j)=0, \label{Approx_WI}
\end{eqnarray}
where $r_i$ are the ratio between the dissipation coefficients $\Upsilon_i$ and Hubble parameter $H$, i.e. $r_i\equiv\Upsilon_i/3H$. As general, it's necessary to introduce some parameters to describe the slow-roll
condition in warm inflation
\begin{subequations}\label{slow-roll_parameters}
\begin{eqnarray}
    \varepsilon &=& \frac{1}{16 \pi G}\Big(\frac{V,_\phi}{V}\Big)^2 \ll 1 , \label{varepsilon_multi} \\
    \eta_{ii} &=& \frac{1}{8 \pi G}\frac{V,_{ii}}{V} \ll 1+r_i , \label{eta_multi}
\end{eqnarray}
and
\begin{eqnarray}
    \beta_i &=& \frac{1}{8 \pi G}\frac{ \Upsilon_{i},_i V,_i}{\Upsilon_i V} \ll 1+r_i. \label{beta_multi}
\end{eqnarray}
\end{subequations}
The term $\Upsilon_\phi,_\chi\dot{\phi}/H^2$ is also a parameter much smaller than the unit but a little differs from slow-roll parameter $\beta_i$:
\begin{eqnarray}
    \frac{\Upsilon_\phi,_\chi\dot{\phi}}{H^2}\simeq-\frac{\Upsilon_\phi,_\phi\frac{\partial\phi}{\partial\chi}\dot{\phi}}{8\pi G V \Upsilon_\phi(1+r_\phi)}\frac{\Upsilon_\phi}{H}=-\frac{3\beta_1 r_1 \tan\theta}{1+r_1},\label{beta_12}
\end{eqnarray}
where $\tan\theta\equiv\dot{\phi}/\dot{\chi}$ and it will be seen as below that $\theta$ is not an independent variable but a constant relaying on dissipative coefficients $\Upsilon_i$.
Define a new variable $z\equiv k/aH$, thus the partial derivative with respect to cosmic time is equivalent to
\begin{eqnarray}
\frac{\partial}{\partial t}=-(1-\varepsilon)zH\frac{\partial}{\partial z}. \label{partial_z}
\end{eqnarray}
With Eqs. (\ref{slow-roll_parameters})$\sim$(\ref{partial_z}), the multi-dimensional Langevin equations (\ref{EOM_pertubed}) turn into
\begin{widetext}
\begin{subequations}\label{EOM_pertubed_z}
\begin{eqnarray}
    \delta\phi_1''+\frac{1}{z}\Big(1-2\frac{2+3r_1-3\varepsilon}{2}\Big)\delta\phi_1'+\Big(1+\frac{3(\eta_1+\beta_1r_1/(1+r_1))}{z^2}\Big)\delta\phi_1
    =\frac{1}{z^2H^2}\xi_1-\frac{3}{z^2}\tilde{\eta}_{12}\delta\phi_2,\label{EOM_background_z_phi} \\
    \delta\phi_2''+\frac{1}{z}\Big(1-2\frac{2+3r_2-3\varepsilon}{2}\Big)\delta\phi_2'+\Big(1+\frac{3(\eta_2+\beta_2r_2/(1+r_2))}{z^2}\Big)\delta\phi_2
    =\frac{1}{z^2H^2}\xi_2-\frac{3}{z^2}\tilde{\eta}_{21}\delta\phi_1,\label{EOM_background_z_chi}
\end{eqnarray}
\end{subequations}
\end{widetext}
where $\tilde{\eta}_{12}=\eta_{12}-{\beta_1 r_1 \tan\theta}/(1+r_1)$, $\tilde{\eta}_{21}=\eta_{21}-{\beta_2 r_2 \cot\theta}/(1+r_2)$ and prime $'$ denotes the derivative with respect to $z$. Now using $t=H^{-1}\ln(k/Hz)$ together with
\begin{eqnarray}
    \delta(f(x))=\sum_{\{x_0\}}\frac{\delta(x-x_0)}{|f'(x_0)|}, \label{deltaFunction}
\end{eqnarray}
where $x_0$ are zero points of $f(x)$, the fluctuation-dissipation relation of $\xi_i$ becomes
\begin{eqnarray}
    \langle \xi_i(\textbf{k},z)\xi_j^*(\textbf{k}',z')\rangle=2 Q_{ij} \frac{H^4}{k^3} z^4 \delta^3(\textbf{k}-\textbf{k}')\delta(z-z'). \nonumber\\
    \label{dis_flu_relation_xi_i}
\end{eqnarray}
The correlation matrix $Q_{ij}$ is  symmetric and non-negative.

Applying Green's function method \cite{arfken2013mathematical}, the solution of differential equation (\ref{EOM_background_z_phi}) is
\begin{eqnarray}
    \delta\phi_1(\textbf{k},z)&=&\int_z^\infty\text{d}z' g_{11}(z,z')\frac{1}{z'^2} \nonumber\\
    & &\big(H^{-2}\xi_1(\textbf{k},z')-3\tilde{\eta}_{12}\delta\phi_2(\textbf{k},z')\big), \label{solution_phi}
\end{eqnarray}
where
\begin{eqnarray}
    g_{11}(z,z')=\frac{z^{\alpha_1}}{z'^{\alpha_1}}\frac{1}{2/\pi z'}\big(J_{\nu_1}(z)Y_{\nu_1}(z')\nonumber\\
    \quad\quad\quad-J_{\nu_1}(z')Y_{\nu_1}(z)\big)\quad\text{for}\ z'>z,\label{g_11}
\end{eqnarray}
with
\begin{subequations}
\begin{eqnarray}
   \alpha_1&=&3(1+r_1-\varepsilon)/2, \label{alpha_1} \\ \nu_1&=&\sqrt{\alpha_1^2-\frac{3\beta_1 r_1}{1+r_1}-3\eta_{11}}.  \label{nu_1}
\end{eqnarray}
\end{subequations}
A detailed description of solution (\ref{solution_phi}) is given in Ref. \cite{PhysRevD.97.063516}. Similarly, the solution of differential equation (\ref{EOM_background_z_chi}) is obtained in the same way:
\begin{eqnarray}
    \delta\phi_2(\textbf{k},z)&=&\int_z^\infty\text{d}z' g_{22}(z,z')\frac{1}{z'^2} \nonumber\\
    & &\big(H^{-2}\xi_2(\textbf{k},z')-3\tilde{\eta}_{21}\delta\phi_1(\textbf{k},z')\big). \label{solution_chi}
\end{eqnarray}
Inserting Eq. (\ref{solution_chi}) into Eq. (\ref{solution_phi}) leads to a Volterra integral equation of the second kind for $\delta\phi_1(\textbf{k},z)$
\begin{eqnarray}
    & &\delta\phi_1(\textbf{k},z)\nonumber \\
    &=&\int_z^\infty\text{d}z' {H^{-2}}\big[\tilde{g}_{11}(z,z') {\xi_1(\textbf{k},z')}
    -3\tilde{\eta}_{12}\tilde{g}_{12}(z,z'){\xi_2(\textbf{k},z')}\big]\nonumber \\& &-9\tilde{\eta}_{12}\tilde{\eta}_{21}{H^{-2}}
    \int_z^\infty\text{d}z' \tilde{g}_{12}(z,z')\delta\phi_1(\textbf{k},z'), \label{solution_int_phi}
\end{eqnarray}
where $\tilde{g}_{11}(z,z')={g}_{11}(z,z')/z'^2$ and
\begin{eqnarray}
    \tilde{g}_{12}(z,z')=\int_z^{z'}\text{d}z'' \tilde{g}_{11}(z,z'')\tilde{g}_{22}(z'',z'). \label{g_12}
\end{eqnarray}
The solution of $\delta\phi_2(\textbf{k},z)$ is an analogue of the formula above. The analytical solution of integral equation (\ref{solution_int_phi}) is a chronological exponential form \cite{srednicki2007quantum, FOX1978179}, which, however, may be not helpful to the calculation on power spectra.
Then the spectra could be obtained in a more straightforward way. Define the autocorrelation matrix ${P}_{ij}(z,z')\equiv\langle\delta\phi_i(\textbf{k},z)\delta\phi_j^*(\textbf{k}',z')\rangle$ and the Green's Function matrix \begin{eqnarray}
    \textbf{g}(z,z')=\left( \begin{array}{ccc}
    g_{11}(z,z') & -3\tilde{\eta}_{12}g_{12}(z,z') \\
    -3\tilde{\eta}_{21}g_{21}(z,z') & g_{22}(z,z')
\end{array}
\right ).\label{GF_matrix}
\end{eqnarray}
If consider only second order of $\tilde{\eta}_{ij}$, the autocorrelation matrix simplifies to
\begin{eqnarray}
    {P}_{ij}& &(z,z)=\langle\tilde{\xi}_i(\textbf{k},z)\tilde{\xi}_j^*(\textbf{k},z)\rangle\nonumber\\
    & &-18\tilde{\eta}_{12}\tilde{\eta}_{21}H^{-2} \int_z^\infty \text{d}z' g_{12}(z,z') P_{ij}(z,z').  \label{autocorrelation_matrix}
\end{eqnarray}
In the equation above, for convenience, we have introduced a new stochastic variable:
\begin{eqnarray}
    \tilde{\xi}_i(\textbf{k},z)=\int_z^\infty\text{d}z' {H^{-2}}\tilde{g}_{ij}(z,z') \xi_j(\textbf{k},z'),\label{xi_tilde}
\end{eqnarray}
with sum on $j$. Obviously, Eq. (\ref{autocorrelation_matrix}) is another integral equation. But, fortunately, only when the second order of $\tilde{\eta}_{ij}$ has a significant impact on the total spectra that we need to solve this complete equation, for the leading order of spectra is just the zero order of $\tilde{\eta}_{ij}$.

With the discussion above, the autocorrelation matrix is written as
\begin{eqnarray}
    P_{ij}(&z&)=P_{ij}^{(0)}(z,z)+P_{ij}^{(1)}(z,z)+\mathcal{O}(\tilde{\eta}^2)\nonumber\\
    &=&\langle\tilde{\xi}_i(\textbf{k},z)\tilde{\xi}_j^*(\textbf{k},z)\rangle+\mathcal{O}(\tilde{\eta}^2)\nonumber\\
    &=&H^{-4}\int_z^\infty\text{d}z' \int_z^\infty\text{d}z'' [\tilde{\textbf{g}}(z,z')]_{ik}\times \nonumber\\
    & &\quad\langle\xi_k(\textbf{k},z')\tilde{\xi}_l^*(\textbf{k}',z'')\rangle[\tilde{\textbf{g}}^\dagger(z,z'')]_{lj}+\mathcal{O}(\tilde{\eta}^2)\nonumber\\
    &=&2\delta^{3}(\textbf{k}-\textbf{k}')\int_z^\infty\text{d}z' \int_z^\infty\text{d}z'' \nonumber\\
    & &\quad[\textbf{g}(z,z')\textbf{Q}\textbf{g}^\dagger(z,z'')]_{ij}\delta(z'-z'')+\mathcal{O}(\tilde{\eta}^2)\nonumber\\
    &=&2\delta^{3}(\textbf{k}-\textbf{k}')\int_z^\infty\text{d}z' [\textbf{g}(z,z')\textbf{Q}\textbf{g}^\dagger(z,z')]_{ij}+\mathcal{O}(\tilde{\eta}^2). \nonumber\\ \label{autocorrelation_matrix_1}
\end{eqnarray}
According to the fluctuation-dissipation relation, the correlation matrix $Q_{ij}$, generally, consists of dissipative coefficients $\Upsilon_i$ in the form of a diagonal matrix. The correlation matrix, as the most general type, exhibits
\begin{eqnarray}
    \textbf{Q}=\left(\begin{matrix} (2\pi)^3k_\text{B}T \Upsilon_1 & 0 \\
    0 & (2\pi)^3k_\text{B}T \Upsilon_2 \end{matrix} \right),\label{correlation_matrix}
\end{eqnarray}
Then, the autocorrelation matrix shows an integral form
\begin{widetext}
\begin{eqnarray}
    \textbf{P}(z)=\frac{6(2\pi)^3H^2}{k^3}\frac{k_\text{B}T}{H}\delta^3(\textbf{k}-\textbf{k}')\int_z^\infty\text{d}z' 
    \left(\begin{matrix}
    r_1 \big[g_{11}\big]^2+\mathcal{O}(\tilde{\eta}^2) & -3\tilde{\eta}_{12}\big[r_1g_{11}g_{12}+r_2g_{12}g_{22}\big] \\
    -3\tilde{\eta}_{21}\big[r_1g_{21}g_{11}+r_2g_{22}g_{21}\big] & r_2 \big[g_{22}\big]^2+\mathcal{O}(\tilde{\eta}^2)
     \end{matrix} \right)(z,z'). \label{autocorrelation_matrix_2}
\end{eqnarray}
\end{widetext}
where we have used a succinct form $[g_{ij}g_{jk}](z,z')$ to simplify the format of function $g_{ij}(z,z')g_{jk}(z,z')$. Noting that
\begin{eqnarray}
    \int_z^\infty\text{d}z'[g_{11}g_{12}](z,z')=\int_z^\infty\text{d}z'[g_{21}g_{11}](z,z') \label{g112=g211}
\end{eqnarray}
(the proof is in Appendix \ref{App_0}) and autocorrelation matric $\textbf{P}(z)$ is symmetric, one finds $\tilde{\eta}_{12}=\tilde{\eta}_{21}$, which leads to a specific relation about $\theta$ defined in Eq. (\ref{beta_12}):
\begin{eqnarray}
    \theta=\arctan\sqrt{\frac{1+r_1}{1+r_2}\frac{r_2}{r_1}\frac{\beta_2}{\beta_1}}. \label{theta}
\end{eqnarray}
The spectra usually tends to scale-invariant at large scale, i.e., $z\ll 1$, or saying at the end of inflation (for $1/aH\rightarrow 0$). Since $z^\alpha J_\nu(z)\rightarrow 0$ and $z^\alpha N_\nu(z)$ tending to a constant when $z\rightarrow 0$, one can ignore the integration containing Neumann function $Y_\nu(z')$ and needs only to consider the term $[z^\alpha]^2 \int_z^\infty\text{d}z' z'^{2-2\alpha}J^2_\nu(z')$ in the integration $\int_z^\infty\text{d}z'[g_{11}]^2(z,z')$. On the other hand, the integrand $z'^{2-2\alpha}J^2_\nu(z')$ almost equals to zero except a narrow peak distributing at $z\gtrsim1$, so one can treat the lower limit of integral as $0$ when $z<1$, which is presented to illustrate the fact that the multi-fields warm inflation exhibits a scale-invariant spectrum as well. Thus,
\begin{eqnarray}
    & &\int_z^\infty\text{d}z' z'^{2-2\alpha}J^2_\nu(z') \nonumber\\
    &\simeq & \frac{\Gamma(\alpha-1)\Gamma(\nu-\alpha+\frac{3}{2})}{2\sqrt{\pi}\Gamma(\alpha-\frac{1}{2})\Gamma(\alpha+\nu-\frac{1}{2})} \nonumber\\
    &\simeq & \frac{\Gamma(\frac{3}{2}r_1+\frac{1}{2})}{4\Gamma(\frac{3}{2}r_1+1)\Gamma(3r_1+\frac{5}{2})}. \label{integral_1}
\end{eqnarray}
In the equations above, we have used the Schafgeitlin integral formula for double Bessel functions \cite{abramowitz2012handbook, wang1989special} and relevant properties of Gamma functions, together with slow-roll conditions Eqs.(\ref{varepsilon_multi})$\sim$(\ref{beta_multi}). Introduce a new function
\begin{eqnarray}
    &I&(r_1)\equiv \frac{16\pi^3}{2\pi^2}\frac{\pi^2}{4}\cdot 3Hk_\text{B}T \left[z^{\alpha_1}Y_{\nu_1}(z)\right]^2\big|_{z\rightarrow 0}\nonumber\\
    & &\quad\quad\quad\quad\quad\quad\quad\times \int_z^\infty\text{d}z' z'^{2-2\alpha}J^2_\nu(z')\nonumber\\
    &\simeq &24\pi H^2\frac{k_\text{B}T}{H}\frac{2^{3r_1}r_1[\Gamma(\frac{3}{2}r_1+\frac{1}{2})]^3}{(3r_1+1)\Gamma(\frac{3}{2}r_1+1)\Gamma(3r_1+\frac{5}{2})}, \label{I}
\end{eqnarray}
where the approximation of Neumann function
\begin{eqnarray}
   Y_\nu(z) \approx -\frac{\Gamma(\nu)}{\pi}\Big(\frac{2}{z}\Big)^\nu\ (\nu>0,\ z\rightarrow 0^+) \label{Neumann_prop1}
\end{eqnarray}
is applied.

For multi-field inflationary model, the evolution for double scaler field perturbations could be decomposed along two directions: one is parallel to trajectory of the evolution for background fields, which is called adiabatic or curvature component, another is orthogonal to the trajectory, which corresponds to the entropy or isocurvature component \cite{PhysRevD.63.023506}. The essential idea to describe the adiabatic field is to introduce the linear combination of perturbed fields $\delta\phi_i$
\begin{eqnarray}
   \delta\sigma\equiv \sin\theta\delta\phi+\cos\theta\delta\chi=\boldsymbol{\varphi}^\dagger\boldsymbol{\theta}, \label{delta_sigma}
\end{eqnarray}
where
\begin{eqnarray}
   \sin\theta\equiv\frac{\dot{\phi}}{\sqrt{\dot{\phi}^2+\dot{{\chi}^2}}}\ \text{and}\ \boldsymbol{\theta}\equiv(\sin\theta,\cos\theta)^T. \label{sin_theta}
\end{eqnarray}
It is obvious why we define the ratio between $\dot{\phi}$ and $\dot{\chi}$ as a tangent relation in Eq. (\ref{beta_12}).
\begin{figure}
  \centering
  \includegraphics[width=3.5in,height=2.5in]{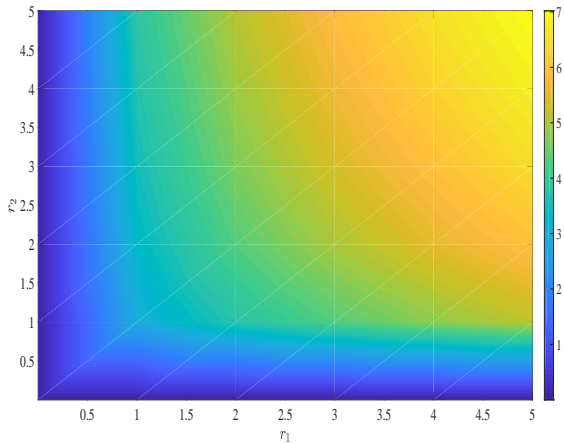}
  \caption{Power spectra for $\mathcal{P}^{(0)}_{\mathcal{R}}$ which is normalized as $\left(\frac{H}{\dot{\sigma}}\right)^2\left(\frac{H}{2\pi}\right)^2=1$ and set to $k_\text{B}T/H\simeq 1/8$ \cite{1475-7516-2016-11-022}.}\label{spectra0}
\end{figure}

In order to quantify the spectra, we define the power spectra of the perturbed fields
\begin{eqnarray}
    \langle\delta\phi_i(\textbf{k},z)\delta\phi_j(\textbf{k}',z)\rangle\equiv \frac{2\pi^2}{k^3}\mathcal{P}_{ij}\delta^3(\textbf{k}-\textbf{k}'). \label{spectra_scale_invariant}
\end{eqnarray}
The spectra of adiabatic field is written as
\begin{eqnarray}
    \mathcal{P}_{\delta\sigma}&=&\mathcal{P}^{(0)}_{\delta\sigma}+\mathcal{P}^{(1)}_{\delta\sigma}+\mathcal{O}(\tilde{\eta}^2)\nonumber\\
    &=&\frac{k^3}{2\pi^2}\int\text{d}^3\textbf{k}'\langle\delta\sigma(\textbf{k},z)\delta\sigma(\textbf{k}',z)\rangle\big|_{z\rightarrow 0}\nonumber\\
    &=&\frac{k^3}{2\pi^2}\int\text{d}^3\textbf{k}' \langle\boldsymbol{\varphi}^\dagger\boldsymbol{\theta}\boldsymbol{\theta}^\dagger\boldsymbol{\varphi}\rangle\big|_{z\rightarrow 0}\nonumber\\
    &=&\frac{k^3}{2\pi^2}\int\text{d}^3\textbf{k}'\text{tr}(\boldsymbol{\theta}\boldsymbol{\theta}^\dagger \textbf{P})\big|_{z\rightarrow 0}. \label{spectra_sigma}
\end{eqnarray}
Inserting Eqs. (\ref{autocorrelation_matrix_2}), (\ref{I}) and (\ref{sin_theta}) into the equation above, we obtain the $i'$th order of the spectra
\begin{eqnarray}
    \mathcal{P}^{(0)}_{\delta\sigma}=I(r_1)\sin^2\theta +I(r_2)\cos^2\theta , \label{spectra_sigma_0}
\end{eqnarray}
and
\begin{eqnarray}
    \mathcal{P}^{(1)}_{\delta\sigma}&=&-6\sin \theta \cos\theta \tilde{\eta}_{12}\times\nonumber\\
    & & \quad \int_0^\infty\text{d}z'\big[r_1g_{11}g_{12}+r_2g_{12}g_{22}\big](z,z'). \label{spectra_sigma_1}
\end{eqnarray}

\begin{figure}
  \centering
  \includegraphics[width=3.5in,height=2.5in]{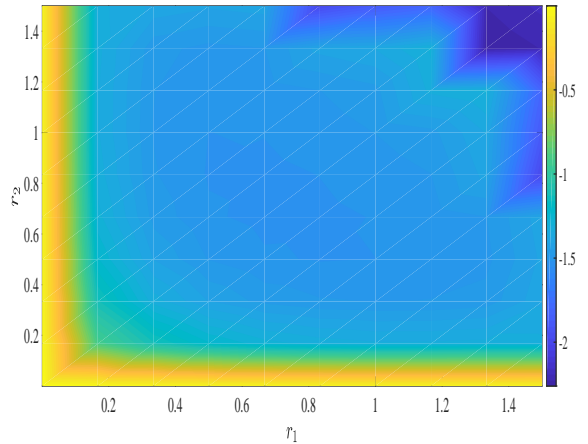}
  \caption{Approximate numerical result of Power spectra for $\mathcal{P}^{(1)}_{\mathcal{R}}$ which is also normalized as $\left(\frac{H}{\dot{\sigma}}\right)^2\left(\frac{H}{2\pi}\right)^2=1$ and set to $k_\text{B}T/H\simeq 1/8$.}\label{spectra1}
\end{figure}

However, the variable $\delta\sigma$ is not physical. The comoving curvature perturbation with spatially flat gauge with observations is given by \cite{PhysRevD.97.063510}
\begin{eqnarray}
    \mathcal{R}=H\frac{\dot{\phi}\delta\phi+\dot{\chi}\delta\chi}{\dot{\phi}^2+\dot{\chi}^2}=H\frac{\delta\sigma}{\dot{\sigma}}. \label{cyrvature_perturbation}
\end{eqnarray}
Finally, we get the spectra of curvature perturbation
\begin{eqnarray}
    \mathcal{P}_\mathcal{R}&=&\frac{k^3}{2\pi^2}\langle\mathcal{R}(\textbf{k},z)\mathcal{R}^*(\textbf{k}',z)\rangle\big|_{z\rightarrow 0}\nonumber\\
     &=&\left(\frac{H}{\dot{\sigma}}\right)^2\mathcal{P}_{\delta\sigma}. \label{spectra_cyrvature_perturbation}
\end{eqnarray}
The analytical result of $\mathcal{P}^{(0)}_{\mathcal{R}}$ is plotted in Fig. \ref{spectra0} while the approximate numerical result of $\mathcal{P}^{(1)}_{\mathcal{R}}$ is plotted in Fig. \ref{spectra1}.

What needs to be pointed out is that the curvature perturbation seems independent on the inflation background, but that's not the case. In statistical physics, the perturbation can not be independent of the background within equilibrium or near equilibrium system. Once a $\theta$ is determined, the trajectory to the evolution of background fields $\dot{\phi}$ and $\dot{\chi}$ is determined as well. Thus the background trajectory leads to the unique perturbation of inflatons $\delta\phi$ and $\delta\chi$. So there exists a one-to-one map between background ($\tan\theta$) and curvature perturbation or isocurvature perturbation. In fact, we can also get a similar result from Eq. (\ref{theta}). The ratio between $\dot{\phi}$ and $\dot{\chi}$ is determined by $r_1$, $r_2$ and alow-roll parameters $\beta_1$ and $\beta_2$, which contains the massage of background trajectory. In a word, the curvature perturbation or isocurvature perturbation depends uniquely on the background trajectory.

\section{\label{entropy_warm_inflation}Entropy of multi-field warm inflation}                                 

As the previous work in multi-fields inflation, the entropy is defined as a linear combination \cite{PhysRevD.63.023506}. But, in this section, we study the entropy in a statistical physical method.

Warm inflationary scenario assume the early Universe is immersed into a thermal bath instead of being cold. Based on such an assumption, the probe on cosmic microwave background shows our Universe is almost on thermal equilibrium state \cite{WMAP}. Thermodynamics shows the entropy, acquiring a maximum value at complete equilibrium, represents a quadratic form in the near equilibrium regime. If the perturbed physical quantities have a phenomenologically multi-dimensional linear Langevin equation representation, the entropy reads \cite{de2013non, ebeling2005statistical}:
\begin{eqnarray}
    & &S(\textbf{k}_p,\delta\phi_i(t)) \nonumber\\
    &=&S_0+\frac{\partial S}{\partial \phi_i}\Big|_{\delta\phi_i=0}\delta\phi_i+\frac{1}{2}\frac{\partial^2 S}{\partial \phi_i \partial \phi_j}\Big|_{\delta\phi_i,_j=0}\delta\phi_i\delta\phi_j \nonumber\\
    &=&S_0-\frac{1}{2}k_\text{B}\delta\phi_i(\textbf{k}_p,t)E_{ij}(t)\delta\phi_j(\textbf{k}_p,t). \label{entropy}
\end{eqnarray}
In the equations above, $S(\textbf{k}_p,\delta\phi_i(t))$ denotes the entropy on physical wave-number $\textbf{k}_p$ and $\textbf{E}$ denotes entropy matrix in terms of symmetric and non-negative form. In the following discussion, we will see the entropy matrix $\textbf{E}$ depends on slow-roll parameters and correlation matrix $\textbf{Q}$ instead of an independent physical variable. The probability of occupying the state $\delta\phi_i(\textbf{k}_p,t)$ is $W(\delta\phi_i(\textbf{k}_p,t))/W_0$, where $W_0$ denotes full equilibrium. Using the definition of entropy $S=k_\text{B}\ln W(\delta\phi_i(\textbf{k}_p,t))$, the distribution satisfies the Boltzmann-Planck formula:
\begin{eqnarray}
    f(\delta\phi_i)=\frac{(\text{det}\textbf{E})^{\frac{1}{2}}}{2\pi}\exp\Big[-\frac{1}{2}\delta\phi_i(t)E_{ij}(t)\delta\phi_j(t)\Big]. \label{distribution}
\end{eqnarray}

Based on slow-roll approximation, the second derivative could be ignored and the multi-dimensional Langevin equations become
\begin{subequations}\label{EOM_pertubed_1}
\begin{eqnarray}
    \dot{\delta\phi_{1}}&+&
    \left(\frac{1}{3(1+r_{1})}\frac{k^{2}}{a^{2}H^{2}}+\frac{\eta_{1}}{1+r_{1}}-\frac{\beta_{1}r_{1}}{(1+r_{1})^{2}}\right)\delta\phi_{1}+ \nonumber\\
    & &\left(\frac{\eta_{12}}{1+r_{1}}-\frac{\beta_{1}r_{1}\tan\theta}{(1+r_{1})^{2}}\right)\delta\phi_{2}=
    \frac{\xi_{1}}{3(1+r_{1})H^2}, \nonumber\\
    \label{EOM_pertubed_linear_phi}\\
    \dot{\delta\phi_{2}}&+&
    \left(\frac{1}{3(1+r_{2})}\frac{k^2}{a^{2}H^{2}}+\frac{\eta_{22}}{1+r_{2}}-\frac{\beta_{2}r_{2}}{(1+r_{2})^{2}}\right)\delta\phi_{2}+ \nonumber\\
    & &\left(\frac{\eta_{12}}{1+r_{2}}-\frac{\beta_{2}r_{2}\cot\theta}{(1+r_2)^2}\right)\delta\phi_{1}
    =\frac{\xi_{2}}{3(1+r_{2})H^2}. \nonumber\\
    \label{EOM_pertubed_linear_chi}
\end{eqnarray}
\end{subequations}
These are just the relations acquired in Eq. (\ref{entropy}) as linear stochastic differential equations. To simplify Eq. (\ref{EOM_pertubed_1}), we introduce a new variable
\begin{eqnarray}
    \textbf{a}(t)=\textbf{R}^{-1}\boldsymbol{\varphi}(t) \label{a}
\end{eqnarray}
with
\begin{eqnarray}
    \textbf{R}=\left(\begin{matrix} \frac{1}{3(1+r_1)} & 0 \\
    0 & \frac{1}{3(1+r_2)} \end{matrix} \right), \label{R}
\end{eqnarray}
and a new matrix
\begin{eqnarray}
{\bf{G}}= \left( \begin{array}{ccc}
\frac{\eta_{11}}{1+r_{1}}-\frac{\beta_{1}r_{1}}{(1+r_{1})^{2}} & \frac{\eta_{12}}{1+r_{1}}-\frac{\beta_{1}r_{1}}{(1+r_{1})^{2}}\tan\theta\\
\frac{\eta_{12}}{1+r_{2}}-\frac{\beta_{2}r_{2}}{(1+r_{2})^{2}}\cot\theta & \frac{\eta_{22}}{1+r_{2}}-\frac{\beta_{2}r_{2}}{(1+r_{2})^{2}}
\end{array}\right ).\nonumber\\
\end{eqnarray}
Thus, the analytical solution to Eq. (\ref{EOM_pertubed_1}) reads
\begin{eqnarray}
    a_{i}(t)=[\text{e}^{-{\textbf{G}}t}]_{ij}a_{j}(0)+\int^{t}_{0}[\text{e}^{-{\textbf{G}}(t-s)}]_{ij}\tilde{\xi}_{j}\text{d}s \nonumber\\
    -z^2\int^{t}_{0}[\textbf{R}e^{-{\textbf{G}}(t-s)}]_{ij}a_{j}(s)\text{d}s  \label{solution_linear_multi}
\end{eqnarray}
with $z\equiv k_p/H=k/aH$. The stochastic force $\tilde{\xi}_i\equiv H^{-2}\xi_i$ also follows the fluctuation-dissipation relation $\langle\tilde{\xi}_i(t)\tilde{\xi}_j^*(t')\rangle=2Q_{ij}a^{-3}(t)\delta(t-t')$, and the exponential map of matrix in Eq. (\ref{solution_linear_multi}) represents $\text{e}^{-\textbf{G}}\equiv\exp[-\textbf{G}]$. As the discussion in Sec. \ref{spectra}, the multi-field warm inflation exhibits a scale-invariant spectra on large scale, which means the autocorrelation matrix of Eq. (\ref{autocorrelation_matrix}) is invariant with time, or statistic physically speaking, variable $a^{3/2}\delta\phi_i$ follows a stationary process. Next we will prove this conclusion via statistic physical method.

\subsection{\label{entropy_large_scale}Entropy on large scale}

A large scale condition means that the parameter $z$ tends to be zero, so the last term in Eq. (\ref{solution_linear_multi}) is neglected. Thus, we simplify it as a tight form:
\begin{eqnarray}
    \dot{{\textbf{a}}}+{\textbf{G}}{\textbf{a}}=\tilde{{\boldsymbol{\xi}}}.  \label{solution_large_scale}
\end{eqnarray}
The solution to differential equation (\ref{solution_large_scale}) reads
\begin{equation}
    {\textbf{a}}(t)=\text{e}^{-{\textbf{G}}t}{\textbf{a}}(0)+\int^{t}_{0}\text{e}^{-{\textbf{G}}(t-s)}{\textbf{a}}(s)\text{d}s. \label{solution_large_scale_1}
\end{equation}
The initial states ${\textbf{a}}(0)$ is determined by the Gaussian distribution of Eq. (\ref{distribution}). The two time autocorrelation matrix $\chi_{ij}(t_1,t_2)$ is introduced as statistical average on two perturbed fields at different times, for $t_1\geqslant t_2$:
\begin{widetext}
\begin{eqnarray}
    \chi_{ij}(t_1&,&t_2)\equiv\{\langle a_{i}(t_{1})a_{j}(t_{2}) \rangle\} \nonumber\\
    &=& [\text{e}^{-{\textbf{G}}t_{1}}]_{ik}[\text{e}^{-{\textbf{G}}t_{2}}]_{jl}\{\langle a_{k}(0)a_{l}(0)\rangle\}
     +\int^{t_{1}}_{0}\text{d}s_{1}\int^{t_{2}}_{0}\text{d}s_{2}[\text{e}^{-{\textbf{G}}(t_{1}-s_{1})}]_{ik}[\text{e}^{-{\textbf{G}}(t_{2}-s_{2})}]_{jl}\langle\tilde{\xi}_{k}(s_{1})\tilde{\xi}_{l}(s_{2})\rangle \nonumber\\
    &=& [\text{e}^{-{\textbf{G}}(t_{1}-t_{2})}]_{ik}[\text{e}^{-{\textbf{G}}t_{2}}{\tilde{\textbf{E}}}^{-1}(0)e^{-{\textbf{G}}^{\dagger}t_{2}}]_{kj}
     + 2[\text{e}^{-{\textbf{G}}(t_{1}-t_{2})}]_{ik}\int^{t_{2}}_{0}\text{d}s[\text{e}^{-{\textbf{G}}(t_{2}-s)}{\textbf{Q}}\text{e}^{-{\textbf{G}}^{\dagger}(t_{2}-s)}\text{e}^{-3s}]_{kj} \nonumber\\
    &=& [\text{e}^{-{\textbf{G}}(t_{1}-t_{2})}]_{ik}[\text{e}^{-\{{\textbf{G}},\cdot\}_{\dagger}t_{2}}{\tilde{\textbf{E}}}^{-1}(0)]_{kj}
    + 2[\text{e}^{-{\textbf{G}}(t_{1}-t_{2})}]_{ik}\int^{t_{2}}_{0}\text{d}s[\text{e}^{-\{{\textbf{G}},\cdot\}_{\dagger}(t_{2}-s)}{\textbf{Q}}\text{e}^{-3s}]_{kj},\label{2t_autoxorrelation_matrix}
\end{eqnarray}
\end{widetext}
In the equations above, $\tilde{\textbf{E}}\equiv\textbf{R}\textbf{E}\textbf{R}$, $\langle\cdots\rangle$ denotes stochastic average and $\{\cdots\}$ denotes the stochastic average on initial state ${\textbf{a}}(0)$. The matrix operator $\{\textbf{G},\cdot\}_\dagger$ is defined by
\begin{eqnarray}
    \{\textbf{G},\textbf{Q}\}_\dagger\equiv\textbf{G}\textbf{Q}+\textbf{Q}\textbf{G}^\dagger.  \label{antisymmetry_operator}
\end{eqnarray}
When applied to double iterated, there exist
\begin{eqnarray}
    \{\textbf{G},\cdot\}_\dagger^2\textbf{Q} &=& \{\textbf{G}, \{\textbf{G},\textbf{Q}\}_\dagger\}_\dagger \nonumber\\
    &=&\textbf{G}^2{\textbf{M}}+2{\textbf{G}}{\textbf{M}}{\textbf{G}}^{\dagger}+{\textbf{M}}({\textbf{G}}^{\dagger})^2.  \label{antisymmetry_operator_2}
\end{eqnarray}
The identity
\begin{eqnarray}
    \text{e}^{-\textbf{G}t}\textbf{Q}\text{e}^{-\textbf{G}^\dagger t}=\text{e}^{-\{\textbf{G},\cdot\}_\dagger t}\textbf{Q}  \label{antisymmetry_operator_exp}
\end{eqnarray}
is also applied in Eq. (\ref{2t_autoxorrelation_matrix}). According to the detailed calculations in Appendix \ref{App_A}. Two time autocorrelation matrix of Eq. (\ref{2t_autoxorrelation_matrix}) is finally written as
\begin{eqnarray}
\{\langle a_{i}(t_{1})&a_{j}(t_{2}) \rangle\}
= [\text{e}^{-{\textbf{G}}(t_{1}-t_{2})}]_{ik}[\text{e}^{-\{{\bf{M}},\cdot\}_{\dagger}t_{2}}\;{\tilde{\textbf{E}}}^{-1}(0)]_{kj} \nonumber\\
&+ [\text{e}^{-{\textbf{G}}(t_{1}-t_{2})}]_{ik}\left[-2\frac{\text{e}^{-3t}}{\lambda-\hat{L}}{\textbf{Q}}+2\frac{\text{e}^{-t\hat{L}}}{\lambda-\hat{L}}{\textbf{Q}}\right]_{kj}, \nonumber \\
\label{2t_autoxorrelation_matrix_final}
\end{eqnarray}
with $\lambda=3$ and $\hat{L}=\{\textbf{G},\cdot\}_\dagger$. Setting $t_1=t_2=t$, the statistical variance matrix of $a_i(t)$ reads
\begin{eqnarray}
& &\left\{ \left\langle a_i(t)a_j(t)\right\rangle \right\}   \nonumber \\
&=&{\tilde{{\bf E}}}_{ij}^{-1}(t)=\chi _{ij}(0)  \nonumber \\
&=&\left[ \text{e}^{-t\hat{L}}{\tilde{{\bf E}}}^{-1}(0)-2\frac{\text{e}^{-3t}%
}{\lambda -\hat{L}}{\bf Q}+2\frac{\text{e}^{-t\hat{L}}}{\lambda -\hat{L}}%
{\bf Q}\right] _{ij}.  \label{variance_ai}
\end{eqnarray}
The unique solution to Eq. (\ref{variance_ai}) is obvious
\begin{equation}
    {\tilde{\textbf{E}}}^{-1}(t)=-2\frac{\text{e}^{-3t}}{\lambda-\hat{L}}{\textbf{Q}}=2\frac{\text{e}^{-3t}}{\hat{L}-\lambda}{\textbf{Q}}, \label{solution_2t_matrix}
\end{equation}
which leads to a relation $E_{ij}(t)=a^3(t)E_{ij}(0)$. When applying $\lambda=0$, Eq. (\ref{solution_2t_matrix}) degenerates to Minkowski condition $\{\textbf{G},\textbf{E}^{-1}\}=2\textbf{Q}$ \cite{FOX1978179, doi:10.1063/1.1693183, doi:10.1063/1.1692878}. Thus, two time autocorrelation matrix of Eq. (\ref{2t_autoxorrelation_matrix}) simplifies to
\begin{eqnarray}
    \{\langle a_{i}(t_1)a_{j}(t_2) \rangle\}=2[\text{e}^{-{\textbf{G}}(t_{1}-t_{2})}]_{ik}\left[\frac{\text{e}^{-3t_2}}{\hat{L}-\lambda}{\textbf{Q}}\right]_{kj}. \label{2t_autocorrelation_matrix_simplified}
\end{eqnarray}
Obviously, two time autocorrelation matrix $\tilde{\chi}_{ij}(t_1-t_2)\equiv a^{3/2}(t_1)a^{3/2}(t_2) \{\langle a_{i}(t_1)a_{j}(t_2) \rangle\}$ exhibits a stationary process which means a stationary expectation variance invariant with time \cite{prabhu2007stochastic}. In other words , the spectra is time invariant on super-horizon scale which is self-consistent with the discussions in Sec. \ref{spectra}. Using the conclusions above, we will get more interesting results on both small and cross-horizon scale.

\subsection{\label{small_scale}Small and cross-horizon scale}

Turn to the discussion on small and cross-horizon scale. The parameter $z$ in solution of Eq. (\ref{solution_linear_multi}) becomes essential under this condition, so the last term containing $z^2$ could not be ignored. We are now looking for the exact analytical solution to Eq. (\ref{EOM_pertubed_1}). Using successive approximation method \cite{jerri1999introduction, tricomi2012integral}, the solution to Eq. (\ref{solution_linear_multi}) decomposes into the terms as follow:
\begin{eqnarray}
    {\textbf{a}}(t)={\textbf{a}}_{0}(t)-z^{2}{\textbf{a}}_{1}(t)+(-z^{2})^{2}{\textbf{a}}_{2}(t)+\cdots, \label{succesive_approximation}
\end{eqnarray}
where
\begin{subequations}\label{a_n}
\begin{eqnarray}
    {\textbf{a}}_{0}(t)&=&\text{e}^{-{\textbf{G}}t}{\textbf{a}}(0)+\int^{t}_{0}\text{e}^{-{\textbf{G}}(t-s)}\tilde{\boldsymbol{\xi}}(s)\text{d}s\equiv{{\textbf{f}}(t)}, \\
    {\textbf{a}}_{1}(t)&=&\int^{t}_{0}\text{d}s\;{\textbf{R}}\;\text{e}^{-{\textbf{G}}(t-s)}{\textbf{a}}_{0}(s), \\
    \cdots \nonumber\\
    {\textbf{a}}_{n}(t)&=&\int^{t}_{0}\text{d}s\;{\textbf{R}}\;\text{e}^{-{\textbf{G}}(t-s)}{\textbf{a}}_{n-1}(s).
\end{eqnarray}
\end{subequations}
After a tedious calculation (see Appendix \ref{App_B}), we get the expression of $\textbf{a}_n(t)$ as
\begin{eqnarray}
    (-z^2)^n{\textbf{a}}_{n}(t) &=& -z^2{\textbf{R}}\int^{t}_{0}ds\frac{(-z^2)^{n-1}}{(n-1)!}[h({\textbf{G}},{\textbf{R}},t-s)]^{n-1}\nonumber\\
    & &\quad\quad\quad\quad \times \text{e}^{-{\textbf{G}}(t-s)}f(s), \label{a_n_solution}
\end{eqnarray}
where
\begin{eqnarray}
    h({\textbf{G}},{\textbf{R}},t-s)=\int_{0}^{t-s}\text{d}s'\text{e}^{-[{\textbf{G}},\cdot]_- s'}{\textbf{R}} \nonumber\\
    =\sum^{\infty}_{n=0}\frac{(t-s)^{n+1}}{(n+1)!}[{\textbf{G}},\cdot]_{-}^{n}{\textbf{R}} \label{h}
\end{eqnarray}
with $[\textbf{G},\cdot]_{-}\textbf{R}\equiv \textbf{G}\textbf{R}-\textbf{R}\textbf{G}$. Sum on $n$ leads to the expression
\begin{eqnarray}
    {\textbf{a}}(t)=\textbf{f}(t)-z^2{\textbf{R}}\int^{t}_{0}\text{d}s\;\text{e}^{-z^{2}h({\textbf{G}},{\textbf{R}},t-s)}e^{-{\textbf{G}}(t-s)}\textbf{f}(s). \nonumber\\ \label{a_n_final}
\end{eqnarray}

\begin{figure}
  \centering
  \includegraphics[width=3.0in,height=2.5in]{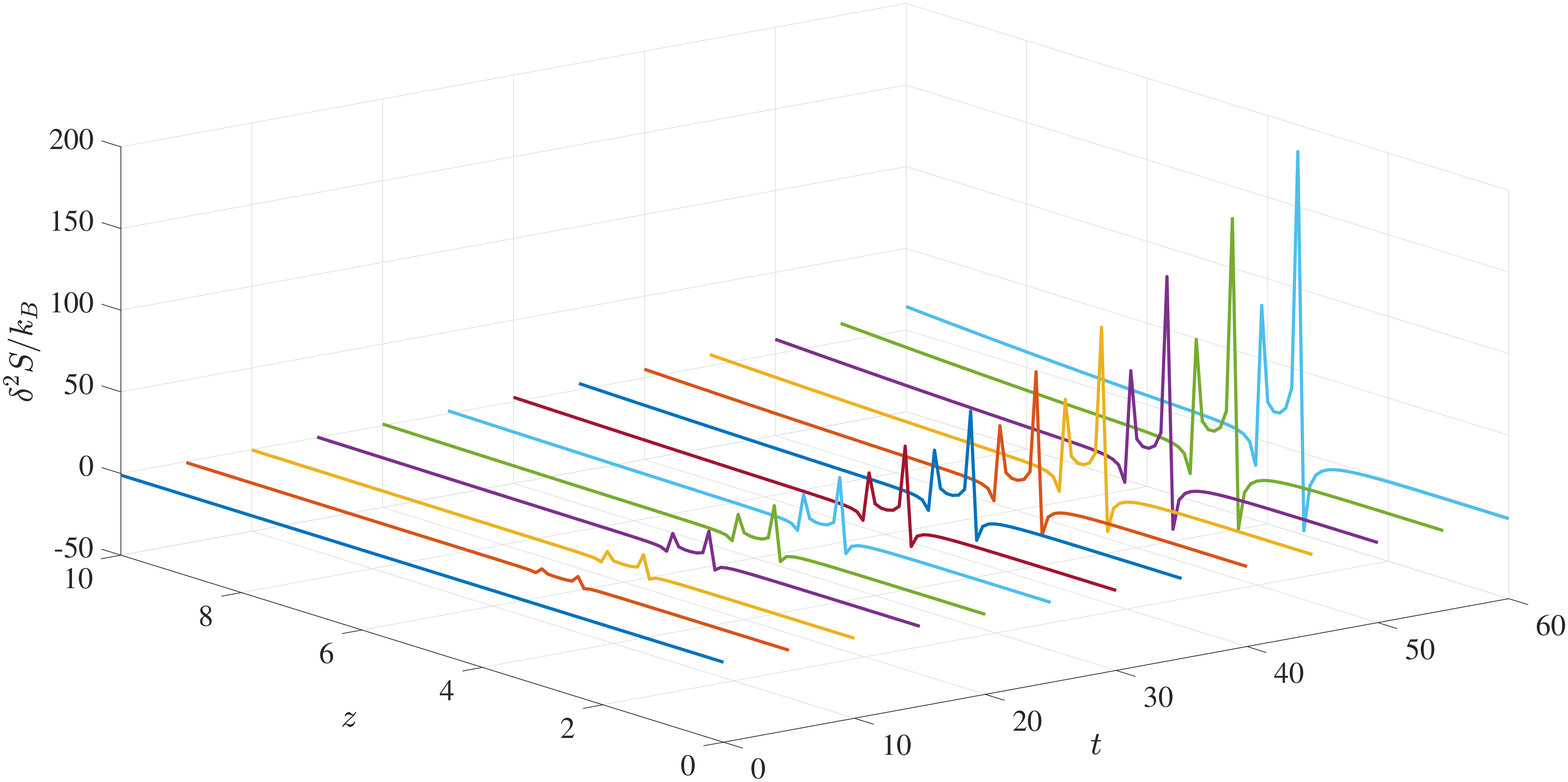}
  \caption{The second order of perturbed entropy $\delta^2S(t)$ as a function of $z$ and $t$. We have set the parameters as $r_1=1.0$, $r_2=0.4$, ${\eta}_{11}/(1+r_1)=0.08$, $\eta_{22}/(1+r_2)=0.05$, ${\eta}_{11}=0.05$, $\beta_1=0.1$ and  $\beta_1=0.05$, which leads to matrix $\textbf{G}=(0.1650\ -0.0251;\;-0.0358\ 0.1194)$. The correlation matrix $\textbf{Q}$ is a diagonal form.}\label{entropy2}
\end{figure}

It employs the variance matrix
\begin{eqnarray}
    \boldsymbol{\Xi}(t)\equiv \langle \textbf{a}(t)\textbf{a}^\dagger(t) \rangle. \label{autocorrelation_matrix_Xi}
\end{eqnarray}
To be convenient on calculation, we decompose matrix $\boldsymbol{\Xi}$ into three components. The first component reads
\begin{eqnarray}
    \boldsymbol{\Xi}_1\equiv\langle{\textbf{f}}(t){\textbf{f}^{\dagger}}(t)\rangle=\frac{2e^{-3t}}{\hat{L}-\lambda}{\textbf{Q}}, \label{Xi_1}
\end{eqnarray}
where we have used the conclusion in Eq. (\ref{2t_autocorrelation_matrix_simplified}). The second is defined as
\begin{eqnarray}
    &&\boldsymbol{\Xi}_{2}(t) \nonumber\\
    &=& \langle(-z^{2}){\textbf{R}}\int^{t}_{0}\text{d}s\;\text{e}^{-z^{2}h({\textbf{G}},{\textbf{R}},t-s)}
    \text{e}^{-{\textbf{G}}(t-s)}\;\textbf{f}(s)\;{\textbf{f}}^{\dagger}(t)\rangle \nonumber\\
    &=& -z^{2}{\textbf{R}}\int^{t}_{0}\text{d}s\;\text{e}^{-z^{2}h({\textbf{G}},{\textbf{R}},t-s)}\text{e}^{-3s} \nonumber\\
    & & \quad\quad\quad\quad\quad\quad \times\text{e}^{-{\textbf{G}}(t-s)}\frac{1}{\hat{L}-\lambda}{\textbf{Q}}\;\text{e}^{-{\textbf{G}}^{\dagger}(t-s)} \nonumber\\
    &=& -z^{2}\text{e}^{-3t}{\textbf{R}}\int^{t}_{0}\text{d}s\;\text{e}^{3s}\text{e}^{-z^{2}h({\textbf{G}},{\textbf{R}},s)} \nonumber\\
    & & \quad\quad\quad\quad\quad\quad \times\text{e}^{-{\textbf{G}}s}\frac{1}{\hat{L}-\lambda}{\textbf{Q}}\;\text{e}^{-{\textbf{G}}^{\dagger}s}. \label{Xi_2}
\end{eqnarray}
While, the third is
\begin{eqnarray}
    \boldsymbol{\Xi}_{3}(t)=(z^{2})^{2}\text{e}^{-3t}{\textbf{R}}\int^{t}_{0}\text{d}s\;\text{e}^{3s}\text{e}^{-z^{2}h({\textbf{G}},{\textbf{R}},s)} \nonumber\\
    \times\text{e}^{-\textbf{G}s}\frac{1}{\hat{L}-\lambda}\textbf{Q}\;\text{e}^{-\textbf{G}^{\dagger}s}\text{e}^{-z^{2}[h(\textbf{G},\textbf{R},s)]^\dagger}\textbf{R}. \label{Xi_3}
\end{eqnarray}
The problems next focus on the integration in Eqs. (\ref{Xi_2}) and (\ref{Xi_3}). However, it's almost impossible to get analytical results, so the numerical results are given for second order of perturbed entropy $\delta^2S(t)$. It employs $\delta^2S(t)$  as (sum on $i$, $j$)
\begin{eqnarray}
    \delta^{2}S(t)
    &=& -\frac{1}{2}k_\text{B}\langle\delta\phi_{i}(t)E_{ij}(t)\delta\phi_{j}(t)\rangle \nonumber\\
    &=& -\frac{1}{2}k_\text{B}\text{tr}({\textbf{E}}(t)\langle\boldsymbol{\varphi}\boldsymbol{\varphi}^{\dagger}(t)\rangle) \nonumber\\
    &=& -\frac{1}{2}k_\text{B}\text{tr}({\textbf{E}}(t)\langle{\textbf{R}}\;{\textbf{a}}(t)\;{\textbf{a}}^{\dagger}(t)\;{\textbf{R}}\rangle) \nonumber\\
    &=& -\frac{1}{2}k_\text{B}\text{tr}(\tilde{\textbf{E}}(t)\;{\boldsymbol{\Xi}}(t)).
\end{eqnarray}
\begin{figure}
  \centering
  \includegraphics[width=3.0in,height=2.5in]{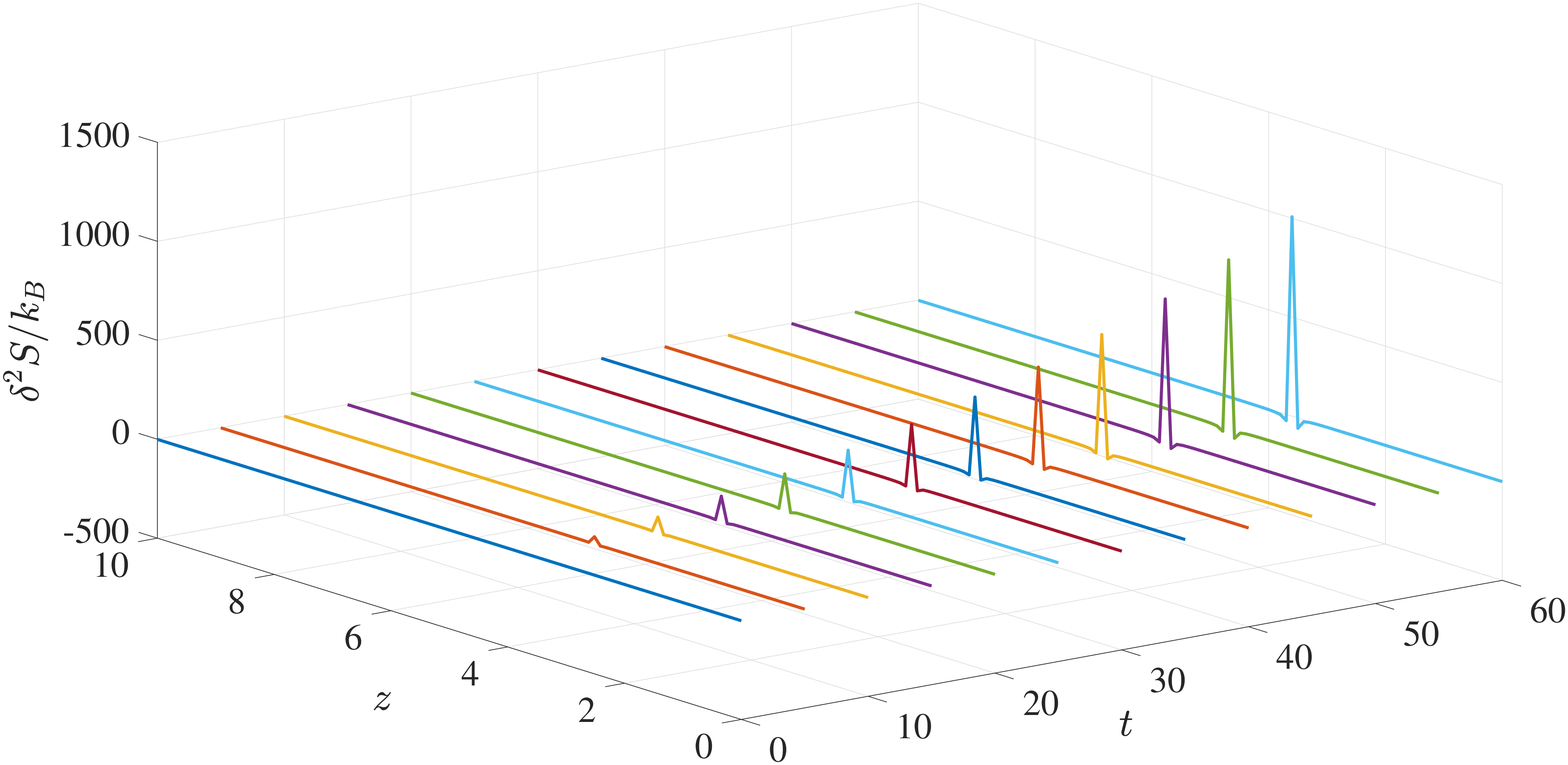}
  \caption{The second order of perturbed entropy $\delta^2S(t)$ as a function of $z$ and $t$. We have set the parameters as $r_1=r_2=0.5$, ${\eta}_{11}/(1+r_1)=0.08$, ${\eta}_{22}/(1+r_2)=0.05$, ${\eta}_{11}=0.05$, $\beta_1=0.1$ and  $\beta_1=0.05$, which leads to matrix $\textbf{G}=(0.1733\ -0.0271;\;-0.0271\ 0.1167)$. The correlation matrix $\textbf{Q}$ is in the form of diagonalization same as the one in Eq. (\ref{correlation_matrix}).}\label{entropy1}
\end{figure}
In Fig. \ref{entropy1}, we plot the a three-dimensional picture to illustrate the relations among second order of perturbed entropy $\delta^2S(t)$, dimensionless scaler variable $z$ and dimensionless cosmic time $t$, which shows several interesting properties. The perturbed entropy almost vanishes at both extremely large and extremely small scale. The former is because the perturbations of fields are freeze-out outside horizon and perturbations no longer increase, which could be treated as an equilibrium state. The latter is because of a sufficient interaction inside horizon, which could be treat as a thermal equilibrium. While, it could be also found a narrow peak locating at $z=3\sqrt{1+r_1}$ approximately. As shown in Eq. (\ref{norm_L}), the value of integrations of Eqs. (\ref{Xi_2}) and (\ref{Xi_3}) rely on the norm of matrix $z^2\textbf{R}+\textbf{G}$. If $z=z_*=3\sqrt{1+r_1}$, it means $\|z^2\textbf{R}+\textbf{G}\|=3$, which exhibits a singularity of the approximate integration
\begin{eqnarray}
    \boldsymbol{\Xi}_{3}(t)&=&(z^{2})^{2}\text{e}^{-3t}{\textbf{R}}\int^{t}_{0}\text{d}s\;\text{e}^{3s}\text{e}^{-\{z^{2}h({\textbf{G}},{\textbf{R}},s),\cdot\}_\dagger} \nonumber\\
    & &\quad\quad\quad\quad\quad\times\text{e}^{-\{\textbf{G},\cdot\}_\dagger s}\frac{1}{\hat{L}-\lambda}\textbf{Q} \nonumber\\
    &\propto&{\textbf{R}}\int^{t}_{0}\text{d}s\;\text{e}^{3s}\text{e}^{-\{(z^2\textbf{R}+\textbf{G}),\cdot\}_\dagger s}\frac{1}{\hat{L}-\lambda}\textbf{Q}.
    \label{Xi_3_approx}
\end{eqnarray}
So $z_*$ is something like the cross-horizon scale. On the other hand, another factor that has an obvious effect on $\delta^2S(t)$ is the slow-roll parameter $\eta_{12}$. Setting $\eta_{12}>1$ (strong interaction situation), it yields $\|z^2\textbf{R}+\textbf{G}\|>3$ no matter how $z$ changes, which shows such evolving curves that the shape peak almost vanishes for any $z$ (see Fig. \ref{entropy3}). This effect is owning to the strong interaction that leads to a obvious departure from slow-roll condition, which, in other words, declares slow-roll condition brings to the generations of entropy. The detailed data also illustrate the increasing of perturbed entropy $\delta^2S(t)$ with time $t$ for any $z$, which follows the second law of thermodynamics. In Fig. \ref{entropy2}, we shows a different situation for $r_1=1.0$, $r_2=0.4$. Two peaks generate at $z_1=3\sqrt{1+r_1}$ and $z_2=3\sqrt{1+r_2}$ respectively and a weak oscillation exists between $z_1$ and $z_2$.

\begin{figure}
  \centering
  \includegraphics[width=3.0in,height=2.5in]{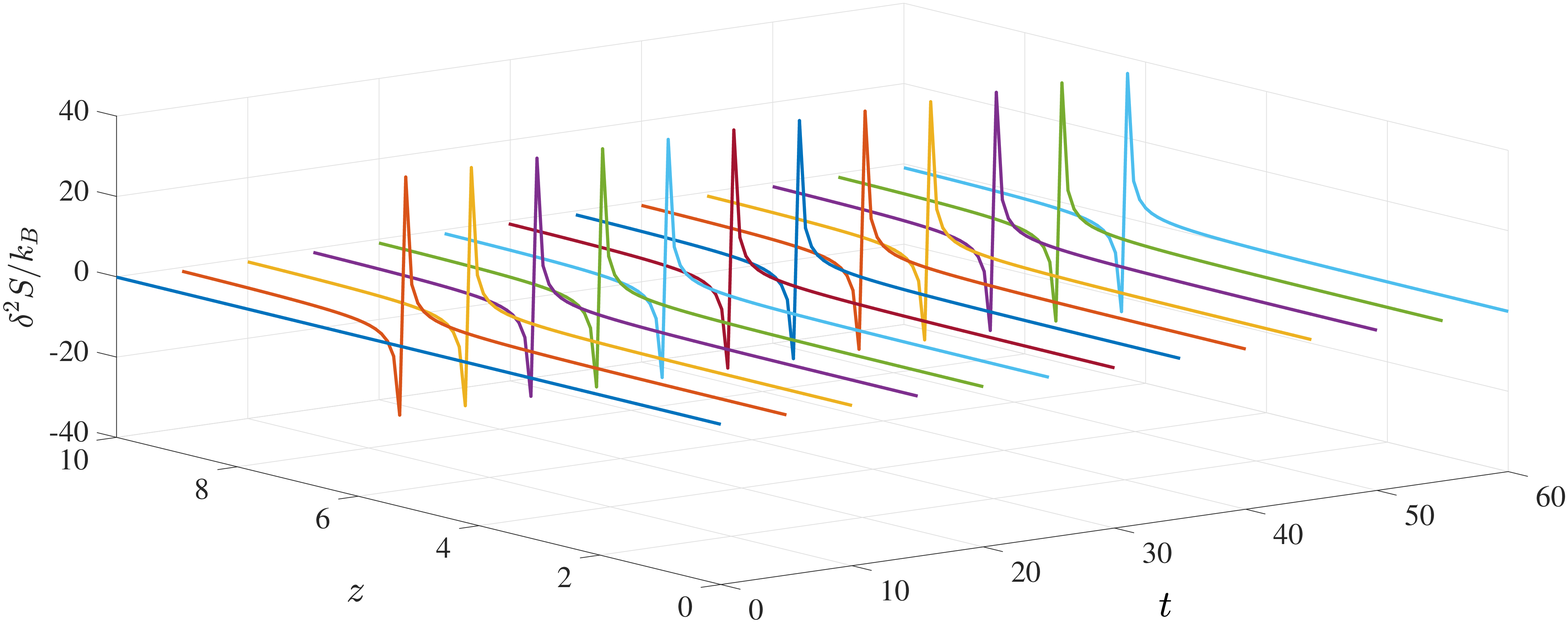}
  \caption{The second order of perturbed entropy $\delta^2S(t)$ with parameters the same as in Fig. \ref{entropy1} except a strong interaction parameter $\eta_{12}=3$. The amplitude is much smaller and there is a oscillation at horizon scale.There is a similar situation with $r_1\neq r_2$.}\label{entropy3}
\end{figure}

\section{\label{conclusion}Conclusion}

In this paper, we have studied the two-field warm inflationary scenario with canonical condition described by a many-dimensional linear stochastic differential equations for it allows an analytical solution. Based on such a model, we calculate its power spectrum and entropy.

First, in Sec. \ref{spectra}, we have calculated analytically the power spectra on super-horizon scale. The perturbed fields is obtained as a form of Volterra integral equation of the second kind. Thus the power spectrum is written as a integral equation as well, but we don't need to solve the complete equation if only considering the leading order and the first order of slow-roll parameter $\eta_{12}$. Generally, the variance of curvature perturbation is introduced as a combination of two perturbed fields whose coefficients are dependent on the background fields. The symmetry of spectra for two perturbed fields leads to a relation between the coefficients and the background fields, or the evolutions of the background fields rely on the dissipative coefficients which is distinguished from those in cold multi-field inflation.

Second, in Sec. \ref{entropy_warm_inflation}, we have calculated analytically the entropy of warm multi-fields inflation on super-horizon scale and sub-horizon scale, as well as cross-horizon scale. In the previous research, the entropy is defined as a linear combination of perturbed fields analogous with the curvature perturbation, but in this paper we introduce it via a symmetric and non-negative matrix which is called entropy matrix. This method has been widely studied in thermodynamics and statistical physics. Similar with the result in flat space-time, the entropy is not an independent parameter but relying on the slow-roll matrix $\textbf{G}$ and correlation matrix $\textbf{Q}$ which is called dissipative-fluctuational relation on super-horizon scale. By solving an integral equation, we obtain analytically the second order of perturbed entropy $\delta^2S(t)$. Then, via numerical analysis, we find several interesting properties of $\delta^2S(t)$. The perturbed entropy almost vanishes at both super-horizon and sub-horizon scale with some appropriate slow-roll parameters. While narrow peaks generates at specific scale interpreted as cross-horizon scale. In addition, the increasing of entropy shows the second law of thermodynamics is followed.

Finally, an extended question, which mentioned at the beginning of this paper, would be investigated how it will be if we extend the multi-field warm inflationary scenario to a noncanonical condition. This question deserves  a further researching.

\begin{acknowledgments}
This work was supported by the National Natural Science Foundation of China (Grants No. 11575270, No. 11175019, No. 11235003).
\end{acknowledgments}

\appendix
\section{\label{App_0}Proof to Eq. (\ref{g112=g211})}
We can prove this equation in such method below. Setting $\hat{L}(\lambda)$ is a differential operator with parameter $\lambda$, the Green's functions $G_1(z,z')$ and $G_2(z,z')$ satisfy
\begin{eqnarray}
    \hat{L}_z(\lambda_1)G_1(z,z')=\delta(z-z'),\\
    \hat{L}_z(\lambda_2)G_2(z,z')=\delta(z-z').
\end{eqnarray}
Consider these functions:
\begin{eqnarray}
    \hat{L}_{z'}(\lambda_1)f(z')=G_2(z',\tilde{z}),\label{F1}\\
    \hat{L}_{z'}(\lambda_2)f'(z')=G_1(z',\tilde{z}),\label{F2}
\end{eqnarray}
where $\tilde{z}$ is an arbitrary with $\tilde{z}>z'$.
The solutions to Eqs. (\ref{F1}) and (\ref{F2}) read
\begin{eqnarray}
    f(z')=\int_z^{z'}\text{d}z''G_1(z,z'')G_2(z'',\tilde{z}),\label{solution_F1}\\
    f'(z')=\int_z^{z'}\text{d}z''G_2(z,z'')G_1(z'',\tilde{z}).\label{solution_F2}
\end{eqnarray}
Apply the operator $\hat{L}_{z'}(\lambda_2)$ on the both sides of Eq. (\ref{F1}) and $\hat{L}_{z'}(\lambda_1)$ on Eq. (\ref{F2}), then we have
\begin{eqnarray}
    \hat{L}_z(\lambda_2)\hat{L}_z(\lambda_1)f(z)=\delta(z-z'),\\
    \hat{L}_z(\lambda_1)\hat{L}_z(\lambda_2)f'(z)=\delta(z-z').
\end{eqnarray}
$\hat{L}_{z'}(\lambda)$ is a differential operator of the Bessel's type, there obviously exist
\begin{eqnarray}
    \hat{L}_z(\lambda_1)\hat{L}_z(\lambda_2)g(z)=\hat{L}_z(\lambda_2)\hat{L}_z(\lambda_1)g(z),
\end{eqnarray}
for any $g(x)\in C^\infty((0,+\infty))$. According to the uniqueness theorem of the solution, it's obviously $f(z')=f'(z')$. Finally set $\tilde{z}\rightarrow z'$, thus we prove the formula in Eq. (\ref{g112=g211}):
\begin{eqnarray}
    g_{12}(z,z')=g_{21}(z,z').
\end{eqnarray}

\section{\label{App_A}Proof to Eq. (\ref{2t_autoxorrelation_matrix_final})}

Let's start from a theorem widely used in functional analysis and spectral theory of linear operator \cite{bachman2012functional, m¨¹ller2013spectral}:
\begin{theorem} \label{inverse_theorem}
Assume linear operator $T$ is a map from a Banach space $X$ to $X$ itself, if the norm of the operator $\|T\|<1$, there exist
\begin{eqnarray}
    \frac{1}{I-T}=\sum_{n=0}^\infty T^n.
\end{eqnarray}
\end{theorem}
The integration in Eq. (\ref{2t_autoxorrelation_matrix}) reads
\begin{eqnarray}
    & &\int^{x}_{0}\text{d}s\;\text{e}^{-(x-s)\hat{L}}\text{e}^{-3s}\nonumber\\
    &=& \sum^{\infty}_{n=0}\frac{\hat{L}^n}{n!}\int^{x}_{0}(s-x)^{n}\text{e}^{-\lambda s}\text{d}s \nonumber\\
    &=& \sum^{\infty}_{n=0}\frac{\hat{L}^n}{n!}\int^{x}_{0}x^{n}(\frac{s}{x}-1)^{n}\text{e}^{-\lambda s}\text{d}s \nonumber\\
    &=& \sum^{\infty}_{n=0}\frac{\hat{L}^n}{n!}\int^{1}_{0}x^{n+1}(s'-1)^{n}\text{e}^{-\lambda xs'}\text{d}s' \nonumber\\
    &=& \sum^{\infty}_{n=0}\frac{\hat{L}^n}{n!}(-1)^{n}\int^{0}_{1}x^{n+1}s^{n}\text{e}^{-\lambda x(1-s)}(-\text{d}s) \nonumber\\
    &=& \sum^{\infty}_{n=0}\frac{\hat{L}^n(-1)^{n}}{n!}x^{n+1}\text{e}^{-\lambda x}\int^{1}_{0}s^{n}\text{e}^{\lambda sx}\text{d}s \nonumber\\
    &=& \sum^{\infty}_{n=0}\text{e}^{-\lambda x}\frac{\hat{L}^n(-1)^{n}}{n!}x^{n+1}\frac{n!}{(-\lambda x)^{n+1}}\nonumber\\
    & & \quad\quad \times\left(1-\sum^{n}_{m=0}\frac{(-\lambda x)^{m}}{m!}\text{e}^{\lambda x}\right) \nonumber\\
    &=& \sum^{\infty}_{n=0}-\frac{\hat{L}^n}{\lambda^{n+1}}(\text{e}^{-\lambda x}-\sum^{n}_{m=0}\frac{(-\lambda x)^{m}}{m!}) \nonumber\\
    &=& \sum^{\infty}_{n=0}-\frac{\hat{L}^n}{\lambda^{n+1}}\sum^{\infty}_{m=n+1}\frac{(-\lambda x)^{m}}{m!} \nonumber\\
    &=& \sum^{\infty}_{m=1}-\frac{(-\lambda x)^{m}}{m!}\sum^{m-1}_{n=0}\frac{(\hat{L})^{n}}{\lambda^{n+1}} \nonumber\\
    &=& \sum^{\infty}_{m=1}-\frac{(-\lambda x)^{m}}{m!}\frac{1-(\frac{\hat{L}}{\lambda})^{m}}{\lambda-\hat{L}} \nonumber\\
    &=& \sum^{\infty}_{m=0}-\frac{(-\lambda x)^{m}}{m!}\frac{1-(\frac{\hat{L}}{\lambda})^{m}}{\lambda-\hat{L}} \nonumber\\
    &=& \frac{\text{e}^{-x\hat{L}}}{\lambda-\hat{L}}-\frac{\text{e}^{-\lambda x}}{\lambda-\hat{L}}.  \label{integration}
\end{eqnarray}
In the calculations above, we have applied Theorem \ref{inverse_theorem} and the integration
\begin{eqnarray}
    \int_0^1\;x^n\text{e}^{-ax}\text{d}x = \frac{n!}{a^{n+1}}\Big(1-\sum^{n}_{m=0}\frac{(-\lambda x)^{m}}{m!}\text{e}^{\lambda x}\Big). \label{int_1}
\end{eqnarray}
In the discussions above, it acquires $\|\hat{L}\|<\lambda$, which, in fact, is quite easy to get satisfied:
\begin{eqnarray}
    \|\hat{L}\|&=&\sup_{\textbf{E}\neq \textbf{0}}\frac{\|\textbf{G}\textbf{E}+\textbf{E}\textbf{G}^\dagger\|}{\|\textbf{E}\|}
    \leqslant\frac{\|\textbf{G}\|\|\textbf{E}+\|\textbf{E}\|\|\textbf{G}^\dagger\|}{\|\textbf{E}\|} \nonumber\\
    &=&\|\textbf{G}\|+\|\textbf{G}^\dagger\|=2\max{\lambda_i}<\lambda=3,\label{norm_L}
\end{eqnarray}
where $\lambda_i$ are eigenvalues of matrix $\textbf{G}$.

\section{\label{App_B}Proof to Eq. (\ref{a_n_solution})}

In this Appendix, we calculate the analytical solution to integral equation (\ref{solution_linear_multi}). Based on successive approximation, the expression of $\textbf{a}_n(t)$ in Eq. (\ref{a_n}) reads
\begin{eqnarray}
{\bf a}_n(t) &=&{\bf R}\int_0^t\text{d}t_1\text{e}^{-{\bf G}(t-t_1)}
\nonumber \\
&&\times \int_0^{t_1}\text{d}t_2{\bf R}\text{e}^{-{\bf G}(t_1-t_2)}\cdots
\nonumber \\
&&\times \int_0^{t_{n-1}}\text{d}t_n\text{e}^{-{\bf G}(t_{n-1}-t_n)}{\bf a}%
_0(t_n)  \nonumber \\
&=&(-z^2){\bf R}\int_0^t\text{d}t_1\int_0^{t_1}\text{d}t_2\cdots   \nonumber
\\
&&\times \int_0^{t_{n-1}}\text{d}t_n\text{e}^{-{\bf G}(t-t_1)}{\bf R}\text{e}%
^{{\bf G}(t-t_1)}\text{e}^{-{\bf G}(t-t_2)}  \nonumber \\
&&\times {\bf R}\text{e}^{{\bf G}(t-t_2)}\text{e}^{-{\bf G}(t-t_3)}\cdots
\nonumber \\
&&\times {\bf R}\text{e}^{{\bf G}(t-t_{n-1})}\text{e}^{-{\bf G}(t-t_n)}{\bf %
f}(t_n)  \label{a_n_1}
\end{eqnarray}
Note that the integrand is symmetric on $t_i$ for $1\leqslant i<n$.
Exchanging order of integration, we have
\begin{eqnarray}
&&\int_0^t\text{d}t_1\int_0^{t_1}\text{d}t_2\cdots \int_0^{t_{n-2}}\text{d}%
t_{n-1}\int_0^{t_{n-1}}\text{d}t_n\cdots   \nonumber \\
&=&\int_0^t\text{d}t_1\cdots \int_0^{t_{n-3}}\text{d}t_{n-2}\int_0^{t_{n-2}}%
\text{d}t_n\int_{t_n}^{t_{n-2}}\text{d}t_{n-1}\cdots   \nonumber \\
&=&\int_0^t\text{d}t_1\cdots \int_0^{t_{n-3}}\text{d}t_n\int_{t_n}^{t_{n-3}}%
\text{d}t_{n-2}\int_{t_n}^{t_{n-2}}\text{d}t_{n-1}\cdots   \nonumber \\
&=&\int_0^t\text{d}t_1\cdots \int_0^{t_{n-3}}\text{d}t_n\int_{t_n}^{t_{n-3}}%
\text{d}t_{n-1}\int_{t_n}^{t_{n-1}}\text{d}t_{n-2}\cdots   \nonumber \\
&=&\int_0^t\text{d}t_1\cdots \int_0^{t_{n-4}}\text{d}t_n\int_{t_n}^{t_{n-4}}%
\text{d}t_{n-3}  \nonumber \\
&&\times \int_{t_n}^{t_{n-3}}\text{d}t_{n-1}\int_{t_n}^{t_{n-1}}\text{d}%
t_{n-2}\cdots   \nonumber \\
&=&\int_0^t\text{d}t_1\cdots \int_0^{t_{n-4}}\text{d}t_n\int_{t_n}^{t_{n-4}}%
\text{d}t_{n-1}  \nonumber \\
&&\times \int_{t_n}^{t_{n-1}}\text{d}t_{n-2}\int_{t_n}^{t_{n-2}}\text{d}%
t_{n-3}\cdots   \nonumber \\
&=&\cdots   \nonumber \\
&=&\int_0^t\text{d}t_n\int_{t_n}^t\text{d}t_{n-1}\int_{t_n}^{t_{n-1}}\text{d}%
t_{n-2}\cdots \int_{t_n}^{t_2}\text{d}t_1\cdots .  \label{a_n_2}
\end{eqnarray}
With Eqs. (\ref{a_n_1}) and (\ref{a_n_2}), ${\bf a}_n(t)$ is simplified to
\begin{eqnarray}
&&(-z^2)^n{\bf a}_n(t)  \nonumber \\
&=&(-z^2)^n{\bf R}\int_0^t\text{d}s\int_s^t\text{d}t_{n-1}\cdots \int_s^{t_2}%
\text{d}t_1  \nonumber \\
&&\times \left[ \text{e}^{-{\bf G}(t-t_1)}{\bf R}\text{e}^{{\bf G}%
(t-t_1)}\right] \cdots   \nonumber \\
&&\times \left[ \text{e}^{-{\bf G}(t-t_{n-1})}{\bf R}\text{e}^{{\bf G}%
(t-t_{n-1})}\right] \text{e}^{-{\bf G}(t-s)}{\bf f}(s)  \nonumber \\
&=&(-z^2)^n{\bf R}\int_0^t\text{d}s\int_s^t\text{d}t_{n-1}\int_s^t\text{d}%
t_{n-2}\cdots   \nonumber \\
&&\times \int_s^t\text{d}t_1\frac 1{(n-1)!}\left[ \text{e}^{-{\bf G}(t-t_1)}%
{\bf R}\text{e}^{{\bf G}(t-t_1)}\right] \cdots   \nonumber \\
&&\times \left[ \text{e}^{-{\bf G}(t-t_{n-1})}{\bf R}\text{e}^{{\bf G}%
(t-t_{n-1})}\right] \text{e}^{-{\bf G}(t-s)}{\bf f}(s)  \nonumber \\
&=&(-z^2){\bf R}\int_0^t\text{d}s\frac{(-z^2)^{n-1}}{(n-1)!}  \nonumber \\
&&\times \left[ \int_s^t\text{d}s^{\prime }\text{e}^{-{\bf G}(t-s^{\prime })}%
{\bf R}\text{e}^{{\bf G}(t-s^{\prime })}\right] ^{n-1}\text{e}^{-{\bf G}%
(t-s)}{\bf f}(s).  \nonumber \\
\label{a_n_3}
\end{eqnarray}
Further more
\begin{eqnarray*}
h({\bf G},{\bf R},t-s) &=&\int_s^t\text{d}s^{\prime }\text{e}^{-{\bf {G}}%
(t-s^{\prime })}{\bf {R}}\text{e}^{{\bf {G}}(t-s^{\prime })} \\
&=&\int_0^{t-s}\text{d}s^{\prime }\sum_{n=0}^\infty \text{e}^{[{\bf G},\cdot
]_{-}s^{\prime }}{\bf R} \\
&=&\sum_{n=0}^\infty \frac{(t-s)^{n+1}}{(n+1)!}[{\bf G},\cdot ]_{-}^n{\bf R},
\end{eqnarray*}
where we have used the relation
\begin{eqnarray*}
\text{e}^{{\bf A}}{\bf B}\text{e}^{-{\bf A}} &=&\text{e}^{[{\bf A},\cdot
]_{-}s^{\prime }}{\bf B} \\
&=&\sum_{n=0}^\infty \frac{(t-s)^{n+1}}{(n+1)!}[{\bf A},\cdot ]_{-}^n{\bf B}
\end{eqnarray*}
with $[{\bf G},\cdot ]_{-}{\bf R}\equiv {\bf G}{\bf R}-{\bf R}{\bf G}$.
According to Eq. (\ref{succesive_approximation}), finally, the solution to
Eq. (\ref{solution_linear_multi}) is written as
\begin{eqnarray}
{\bf a}(t) &=&{\bf f}(t)-z^2{\bf R}\int_0^t\text{d}s\sum_{n=1}^\infty \frac{%
(-z^2)^{n-1}}{(n-1)!}  \nonumber \\
&&\times \left[ \int_s^t\text{d}s^{\prime }\text{e}^{-{\bf G}(t-s^{\prime })}%
{\bf R}\text{e}^{{\bf G}(t-s^{\prime })}\right] ^{n-1}\text{e}^{-{\bf G}%
(t-s)}{\bf f}(s)  \nonumber \\
&=&{\bf f}(t)-z^2{\bf R}\int_0^t\text{e}^{-z^2h({\bf R},{\bf G},t-s)}\text{e}%
^{-{\bf G}(t-s)}{\bf f}(s)\text{d}s.  \nonumber \\
\label{a_final}
\end{eqnarray}


\end{document}